\documentclass[texture,10pt]{article}
\usepackage{epsfig}
\def\beq{\begin{equation}}
\def\eeq{\end{equation}}
\def\ben{\begin{eqnarray}}
\def\een{\end{eqnarray}}
\def\bea{\begin{array}}
\def\eea{\end{array}}

\begin{document}

\baselineskip=18pt
\begin{flushright}
Prairie View A \& M, HEP-9-97\\
September 1997 \\

\end{flushright}

\vskip.1in

\begin{center}
{\Large\bf Comments on the review of CP Violation \\
in the Particle Data Book}
\vskip .3in

{\bf Dan-Di Wu}\footnote{ $~$E-mail:  wu@hp75.pvamu.edu or
danwu@physics.rice.edu }
\vskip.1in

{\sl HEP, Prairie View A\&M University, Prairie View, TX 77446-0355, USA}
\end{center}
\vskip.5in

{\sl Rephasing invariant formalism for the $K^0-\bar 
K^0$ system is recommended for the 
Particle Data Group. This formalism is accurate and prevents possible errors
in generalizing the formalism to other systems, as well as in using
CKM matrices not included in the Particle Data Book.}

\vskip.4in

The 1998 Particle Data Book (PDB) is already available on the Web site. 
In this brief note I will comment on an essential part
of  ``CP Violation" (referred as CPVW) prepared by L.
Wolfenstein. 
Compared with the old version\cite{Wol1} by Wolfenstein and Trippe,
a crucial differentiation
is made, for the study 
of the $K^0$ system, between the parameter $\tilde\epsilon$
 in (12.3) (that is phase convention dependent)
and $\epsilon$ in (12.5a) (that is phase convention independent).
 However a further improvement  may  
help the CLEO group to avoid a mistake in assuming, for the $B_d$ system,
a counterpart of Eq (12.5d),
 $Re\epsilon_{_{B_d}}
\approx\frac{1}{4}a_{ll}$\cite{CLEO}, which is still in the new version of
the PDB. All three recommended forms of the CKM   matrix (RCKM)\cite{CKM}
 in the Particle Data Book provide the same $
a_{ll}$ of order 10$^{-4}$.
But  they give large and different $Re\epsilon_{_{B_d}}$.
Indeed,
$\epsilon_{
_{B_d}}=\frac{(1+\sigma)-(1-\sigma)e^{i\theta}}
{(1+\sigma)+(1-\sigma)e^{i\theta}}$, where $\theta={\rm arg}(V_{tb}
V^*_{td})^2$, and $\sigma=\frac{1}{2}\langle B_H|B_L\rangle\sim 10^{-4}$. 
The phase $\theta$ can be arbitrarily changed by changing the phase of the 
b-quark field, for example. Consequently, $\epsilon_{
_{B_d}}$ can be as large as $1/\sigma\sim 10^4$. 
This is an example of how generalizing a formalism for the $K^0$ system
under 
 specific phase conventions (to be discussed later) leads to mistakes
because for the $B_d$ system
the same conventions (such as RCKM)
do not lead to similar formulas.
\vskip.1in
I find Eqs (12.5b and 12.5c) for the $K^0$ system 
in CPVW misleading. For the reader's convenience, I record these equations:
$$
 (12.5b)\hskip1.5in \epsilon=\tilde\epsilon + i(Im A_0/Re A_0),\ \ \ \ \ \ \ 
 \ $$
$$
 (12.5c)\hskip.4in \sqrt{2}\epsilon^\prime=ie^{i{\delta_2-\delta_0)}}
(Re A_2/Re A_0)(Im A_2/Re A_2-Im A_0)/Re A_0).
$$
 Yet they
are numerically correct, if any one of the three CKM phase conventions
recommended (the RCKM) in the PDB is used, 
because for these conventions $\tilde\epsilon$ for the Kaon is
 small. This precondition for the validity of these equations is missing 
 in the CPVW. A CKM phase convention is just an example of phase convention of
 the coupling constants. If I choose 
Re$ A_0 = 0$ by changing the phase of the $s$ quark field, these
equations get zero denominators. Note to 
 keep $CP|K^0\rangle=|\bar K^0\rangle$, which is a requirement in CPVW, 
I can use the 
 phase of the composite wave function.   
 Now the effect of 
 the phase of the $s$ quark will only show in the form of the 
CKM matrix, which then affect the phases of the relevant decay amplitudes and 
 mixing mass and width.
 Consequently, Re$\tilde\epsilon$ may not be small, 
depending on the CKM phase convention adopted. 
New CKM matrices are recommended by Chen and Wu\cite{Chen} and
by Fritzsch and Xing\cite{Xing} with some of those matrices
these formulas  become
wild.  
In addition the sentence after Eq (12.4) of CPVW, ``$A_I$ would 
be real if CP invariance held" is misleading, 
because only  relative phases between different 
quantities count in physics.  Even if the phase difference between
$A_0$ and $A_2$ did not exist, there could still be a phase difference between
$M_{12}$ and $\Gamma_{12}$ in principle, which would contribute to  CP 
noninvariance in the mixing\cite{Wu80}.
 In the following I list phase convention 
independent formulas, then discuss
 convention dependent formulas (two sets for two conventions that 
 appeared in CPVW). 
Since $\epsilon $ is widely used as a phase convention dependent
parameter for all the mixing systems (e.g. $\epsilon_{_{B_d}}$ discussed
above),
 I will replace $\tilde\epsilon$ by
$\epsilon$ and $\epsilon$ by $\epsilon_0$ in my following presentation.
\vskip.1in
1) The accurate formulas for (12.5b) and (12.5c) should read instead (assuming 
CPT invariance):
\beq
 \epsilon_0=(\epsilon{\rm Re}A_0+i{\rm Im}A_0)/
({\rm Re}A_0 +i\epsilon{\rm Im}A_0)
\eeq
\beq
 \epsilon^\prime=({i}/\sqrt{2})e^{i(\delta_2-\delta_0)}
{\rm Im}({A_2}/{A_0}),
\eeq
These equations are simple.
Eq (1) is very easy to 
reproduce, given the definitions\cite{Yang}\cite{
Wolf} of 
the relevant quantities
$$K_S=\frac{1}{\sqrt{2(1+
|\epsilon|^2)}}[(1+\epsilon)K^0+(1-\epsilon)\bar K^0],$$ 
$$K_L=\frac{1}{\sqrt{2(1+
|\epsilon|^2)}}[(1+\epsilon)K^0-(1-\epsilon)\bar K^0].$$ 
Note that $\epsilon$ thus defined is CKM phase convention dependent after 
assigning CP$|K^0\rangle=|\bar K^0\rangle$.
$$\epsilon_0=\frac{A(K_L\rightarrow 2\pi,I=0)}{A(K_S\rightarrow 2\pi,I=0)}=
(1/3)(2\eta_{+-}+\eta_{00}),
$$
The reduction of Eq (2) can be found, for example, in Ref\cite{Wu2}, given
$\epsilon_0$ small experimentally, where $\epsilon^\prime$ is defined as
$$\epsilon^\prime=\frac{1}{\sqrt{2}}\left[\frac{
\langle 2\pi,I=2|K_L\rangle}{\langle 2\pi,I=2|K_S\rangle}-
\epsilon_0\frac{\langle 2\pi,I=2|K_S\rangle}{\langle 2\pi,I=0|K_S\rangle}
\right]=(1/3)(\eta_{+-}-\eta_{00}).  $$
I-spin symmetry is assumed in formulas for $\epsilon_0 $, 
$\epsilon^\prime$.

\vskip.1in
 2)  Under the Wu-Yang phase convention, 
$${\rm Im}A_0=0,$$
Eqs (1 and 2) become respectively:
\beq
\epsilon_0=\epsilon.
\eeq
\beq
\epsilon^\prime=({i}/\sqrt{2})e^{i(\delta_2-\delta_0)}
{{\rm Im}A_2}/{A_0},
\eeq
\vskip.1in
3) However, under the conventions of the 
RCKM matrices, in particular, the
Wolfenstein matrix\cite{Wol}, 
$${\rm Im}A_2=0.$$
$\epsilon$ under this convention 
is also small because $\epsilon_0$ and Im$A_0$ are both small.
 Indeed, 
$$\epsilon=(\epsilon_0{\rm Re}A_0-i{\rm Im}A_0)/({\rm Re}A_0-i\epsilon_0
{\rm Im}A_0).$$
 We therefore have
\beq
\epsilon_0=\epsilon+i({\rm Im}A_0/{\rm Re}A_0),
\eeq
\beq
\epsilon^\prime=-({i}/\sqrt{2})e^{i(\delta_2-\delta_0)}
({A_2}/{|A_0|})({\rm Im}A_0/|A_0|),
\eeq
 At the moment, the Wu-Yang convention is not widely used 
because the specific CKM matrix to realize the 
Wu-Yang convention is difficult to present. The phase convention
dependent parameter $\epsilon$ in both Wu-Yang conventions and
the RCKM conventions are small, however, this is completely an effect of
phase convention, that has nothing to do with the smallness of $\epsilon_0$
and $\epsilon^\prime$. At the extreme,
if one chooses Re$A_0=0$, 
one gets $\epsilon=1/\epsilon_0\sim 10^{3}
$, which tells that the unphysical parameter
$\epsilon $ can become very large by just changing the conventions. 
Therefore $\epsilon$ is not measurable. The 
arbitrariness of the $\epsilon$ parameter has been thoroughly discussed by Wu
\cite{Wu2} and recently by Xing\cite{Xinge}.

\vskip.1in
 Finally, the formula 
$$Re \epsilon_{_{B_d}}\approx\frac{1}{4}a_{ll}$$
 used by the CLEO group will make sense if $\epsilon_{_{B_d}}$ is made
 very small (at the order of $\sigma=\langle B_H|B_L\rangle$)
 by choosing suitable CKM conventions. 
This is realized  by the Chen-Wu matrix discussed in a 
preprint\cite{Wu}\footnote{Also by the Fritzsch-Xing matrix\cite{Xing}.}.
 With this matrix, 
Eq (12.9) of CPVW is simplified because 
$(q_B/p_B)=$Real is intended in this convention. It so happens
 that the matrix that realizes this convention 
 also makes Eq (12.11) of CPVW for the asymmetry in the process
$B_d \rightarrow \psi K_S$\cite{Sanda}
 to become 
$$a_{\psi K_S}={\rm sin}2\delta$$
where $\delta$ is the phase angle in the Chen-Wu 
 matrix. I believe 
this result is very important  to the BaBar and 
the Belle experiments, given the limited scope and accuracy of these
experiments. It upgrades the measurement from getting one of many angles 
and sides of the unitarity triangles (many of these angles and sides are
not measurable or poorly measurable) to getting one of a set of 
 four parameters of the 
CKM matrix.
\vskip.1in
This work is in part supported by an NSF HRD grant and in part by the Center
for Applied Radiation Research (CARR) at Prairie View A\&M University.

\end{document}